\documentclass[conf]{new-aiaa}
\usepackage[utf8]{inputenc}

\usepackage{graphicx}
\usepackage{amsmath}
\usepackage{bm}
\usepackage{hyperref}
\usepackage{algorithm}
\usepackage{algpseudocode}

\newtheorem{theorem}{Theorem}

\title{Closed-Form Solutions to the Fokker-Planck Equation for Orbital Uncertainty Propagation}

\author{José Antonio Rebollo\footnote{PhD Student, Aerospace Engineering Department, University of Seville, 41092, Seville, Spain.} and Rafael Vázquez\footnote{Professor, Aerospace Engineering Department, University of Seville, 41092, Seville, Spain.}}
\affil{Universidad de Sevilla, Seville, Spain}
\author{Claudio Bombardelli\footnote{Professor, Department of Applied Physics for Aeronautical Engineering, Technical University of Madrid, 28040, Madrid, Spain.}}
\affil{Universidad Politécnica de Madrid, Madrid, Spain}

\begin{document}

\maketitle

\begin{abstract}
Non-Gaussian tails dominate collision probability estimates in conjunction assessment, yet capturing them without Monte Carlo sampling is challenging, especially when process noise is included. We present a closed-form, grid-free solution to the Fokker-Planck equation by proving that an exponential-of-quadratic-form ansatz is structurally preserved under advection and diffusion. The probability density function propagates via a compact ODE system, significantly cheaper than Monte Carlo and without spatial discretization. As an application, the method performs orbit uncertainty propagation under stochastic forcing representative of atmospheric drag. Results demonstrate the method faithfully captures non-Gaussian features, asymmetric tails, and stochastic broadening, matching a Monte Carlo benchmark.
\end{abstract}

% ============================================================
\section{Introduction}
% ============================================================

\lettrine{C}{haracterizing} how uncertainty evolves under nonlinear orbital dynamics is fundamental to modern space operations. Conjunction assessment, collision avoidance maneuver planning, re-entry prediction, and Space Domain Awareness (SDA) all depend on knowing not just where a spacecraft is likely to be, but how probable it is to be somewhere unexpected~\cite{Luo2017}. In these applications, the \emph{tails} of the probability distribution are often more consequential than its peak: a conjunction screening decision, for instance, hinges on whether the probability of collision, typically on the order of $10^{-5}$ or smaller, exceeds an actionable threshold~\cite{Akella2000,Patera2001,Chan2008}. Accurately modeling these low-probability tails requires a faithful representation of the full probability density function (PDF), not merely its first two moments.

Current operational methods overwhelmingly rely on Gaussian assumptions propagated through linearized dynamics, e.g., via covariance propagation with state transition matrices~\cite{Maybeck1979}. These approximations systematically misrepresent the tails of the distribution: nonlinear dynamics stretch, curve, and skew an initially Gaussian cloud into distinctly non-Gaussian shapes~\cite{Junkins1996}, while the Gaussian model continues to assign probability mass symmetrically. The practical consequence is that collision probabilities are either over- or under-estimated, both of which carry real operational cost, unnecessary avoidance maneuvers consuming finite fuel, or missed warnings for genuinely dangerous encounters. Extensions such as the unscented transform~\cite{Julier2004} improve moment propagation but remain fundamentally limited to low-order statistics and cannot reconstruct the full PDF shape. A complementary strategy is to reformulate the problem in a coordinate set that renders the dynamics nearly linear, so that the Gaussian assumption remains approximately valid; the Generalized Equinoctial Orbital Elements (GEqOE) have been shown to achieve this for the deterministic propagation case~\cite{Hernando2023}.

More sophisticated approaches exist. Monte Carlo methods can capture the full nonlinear PDF, but require $10^5$--$10^6$ samples to resolve the $10^{-5}$ tail probabilities relevant to conjunction assessment, making them computationally prohibitive for operational timelines~\cite{Luo2017}. Gaussian mixture models~\cite{Horwood2011,Vittaldev2016} and entropy-based methods~\cite{DeMars2013} improve on the single-Gaussian approximation, but require an increasing number of components or basis functions to capture non-Gaussian tails, and the associated computational cost grows accordingly. Grid-based numerical solvers for the Fokker-Planck equation~\cite{Risken1996,Kumar2009} suffer from the curse of dimensionality: a 6D orbital state on even a coarse $50$-point-per-axis grid requires $50^6 \approx 10^{10}$ evaluations per time step. The Probability Transformation Method (PTM), combined with Differential Algebra (DA) and Taylor maps~\cite{Berz1999,Park2006,Valli2013,Armellin2010,Wittig2015}, elegantly solves the deterministic part of the problem (the Liouville equation) by propagating the flow map as a high-order polynomial expansion. However, all physical systems experience stochastic forcing, atmospheric drag fluctuations, unmodeled gravitational perturbations, solar radiation pressure variability, thrust execution errors, and incorporating this process noise into the PTM framework has remained an open challenge. Previous work has addressed stochastic uncertainty in specific applications using DA-based approaches~\cite{Vazquez2017}, while Servadio and Zanetti~\cite{Servadio2020} developed recursive polynomial methods for nonlinear estimation applied to orbital mechanics. Nevertheless, extending the PTM to the full stochastic setting requires evaluating high-dimensional convolution integrals that reintroduce the very curse of dimensionality that DA was designed to circumvent.

This paper presents Taylor Map Diffusion, a framework that closes this gap. We show that the Fokker-Planck equation, including both deterministic advection and stochastic diffusion, admits solutions of a specific structural form: the exponential of a quadratic function of a nonlinear map. To the authors' knowledge, this exponential-of-quadratic-form solution structure is new in the literature: while exponential-quadratic forms are well known for linear systems (where they reduce to Gaussians propagated by the Kalman filter), their preservation under the combined action of nonlinear advection and additive diffusion has not previously been established. The key result is a set of coupled ordinary differential equations (ODEs) governing the time evolution of the map and the precision matrix. These ODEs are derived by direct substitution into the FPE, yielding zero residual: the structural form is preserved rigorously under the combined action of advection and diffusion. In practice, the map is represented as a finite-order Taylor expansion, so the accuracy of the computed PDF is controlled by the truncation order, converging to the theoretical solution as the expansion order increases. The method is grid-free, captures the full non-Gaussian shape of the distribution including its tails, and naturally incorporates process noise. We demonstrate the algorithm on a planar Keplerian orbit problem and validate it against Monte Carlo stochastic simulations.

% ============================================================
\section{The Taylor Diffusion Framework}
% ============================================================

\subsection{Ansatz and Main Result}

Consider a dynamical system subject to additive white Gaussian noise,
\begin{equation}\label{eq:sde}
    dX = f(X)\,dt + \sigma\,dW_t,
\end{equation}
where $X \in \mathbb{R}^n$, $f$ is the deterministic drift, and $\Sigma_0 = \sigma\sigma^\top$ is the diffusion tensor. The probability density $p(x,t)$ satisfies the Fokker-Planck equation (FPE):
\begin{equation}\label{eq:fpe}
    \frac{\partial p}{\partial t} = -\nabla\cdot(f\,p) + \frac{1}{2}\mathrm{Tr}\!\left(\Sigma_0\,\nabla^2 p\right).
\end{equation}
We propose the following ansatz for the density:
\begin{equation}\label{eq:ansatz}
    p(x,t) = \mathcal{N}(t)\,\mu(x,t)\,\exp\!\Big(-F(x,t)^\top Q(t)\,F(x,t)\Big),
\end{equation}
where $F:\mathbb{R}^n\to\mathbb{R}^n$ is a smooth, invertible map with $F(x_0(t),t) = 0$ (centered at the distribution's mode), $Q(t) \succ 0$ is a symmetric positive definite precision matrix, $\mu(x,t)$ is a smooth scalar field accounting for the divergence of the dynamics, and $\mathcal{N}(t)$ is a normalization constant.

The central result of this work is:

\begin{theorem}[Quadratic Form Conservation]\label{thm:main}
Let the initial density be of the form~\eqref{eq:ansatz} where $F(\cdot,t_0)$ is a diffeomorphism with isolated zero at $x_0(t_0)$, $Q(t_0)\succ 0$, and $\mu(x,t_0)$ a smooth scalar volume field. Under additive white Gaussian noise with diagonal covariance $\Sigma_0 = \sigma\sigma^\top$, this structural form is preserved over time, provided $(F,\mu,Q)$ satisfy the coupled system:
\begin{align}
    \frac{\partial F}{\partial t} &= -J_F\,f + \frac{1}{2}\!\left(F\sum_i\alpha_i + Q^{-1}\!\sum_i D_i + \Delta - \Gamma\right), \label{eq:dFdt}\\[4pt]
    \frac{\partial \mu}{\partial t} &= (\nabla\cdot f) - \nabla\mu\cdot f - (\nabla\mu)^\top\Sigma_0\,J_F + \frac{1}{2}\mathrm{Tr}\!\left(\Sigma_0\,\nabla^2\mu\right), \label{eq:dmudt}\\[4pt]
    \frac{dQ}{dt} &= \frac{1}{2}\sum_i J_F^{-\top}\,J_{\Lambda_i}^\top Q\,J_{\Lambda_i}\,J_F^{-1}, \label{eq:dQdt}
\end{align}
where $J_F = \nabla_x F$ is the Jacobian of $F$; $\Lambda_i = \sqrt{\Sigma_{0,i}}\,\nabla F_i$ with $J_{\Lambda_i} = \nabla_x\Lambda_i$; $\Gamma = 2J_F\Sigma_0 J_F^\top QF$; $\Delta_k = \sum_j \Sigma_{0,j}\,\partial^2 F_k/\partial x_j^2$; and $D_i \in \mathbb{R}^n$, $\alpha_i \in \mathbb{R}$ are smooth coupling fields that regularize the diffusion correction at $F = 0$ (see below).
\end{theorem}

The proof proceeds by direct substitution of the ansatz~\eqref{eq:ansatz} into the FPE~\eqref{eq:fpe}. Computing both sides independently, the time derivative via the chain rule on $\mu e^{-\Phi}$ with $\Phi = F^\top QF$, and the right-hand side via the advection and diffusion operators, one verifies that all terms cancel identically when $(F,\mu,Q)$ satisfy the above system. The coupling fields $D_i$ and $\alpha_i$ arise from a key algebraic identity: any perturbation of a quadratic form $F^\top QF$ by a small symmetric correction can be absorbed into a smooth deformation of the underlying map and metric, provided $D_i$ and $\alpha_i$ are chosen to remove the apparent singularity at $F=0$. Specifically:
\begin{equation}
    D_i = J_F^{-\top} J_{\Lambda_i}^\top Q\,\Lambda_i(x_0), \qquad
    \alpha_i = \frac{\Lambda_i^\top Q\Lambda_i - F^\top J_F^{-\top} J_{\Lambda_i}^\top Q J_{\Lambda_i} J_F^{-1} F - D_i^\top F - C_i}{F^\top QF},
\end{equation}
where $C_i = \Lambda_i(x_0)^\top Q\Lambda_i(x_0)$ is a scalar constant ensuring $\lim_{x\to x_0}\alpha_i = 0$.

\smallskip\noindent\textbf{Remark (theoretical vs.\ numerical).} Theorem~\ref{thm:main} establishes that the structural form~\eqref{eq:ansatz} is closed under the FPE dynamics: no approximation is introduced in the evolution equations~\eqref{eq:dFdt}--\eqref{eq:dQdt}. In a numerical implementation, $F$ is represented as a finite-order Taylor expansion (e.g., up to quadratic terms). This truncation is the sole source of approximation: its effect diminishes as expansion order increases, and for linear dynamics the first-order representation is exact, recovering the result described in the following remark.

\smallskip\noindent\textbf{Remark (classical covariance propagation).}
The linear case provides the clearest illustration of the framework's structure and its generalization. For linear dynamics $f(x) = Ax$, the Fokker-Planck equation admits an exact Gaussian solution, where the PDF is fully characterized at all times by its mean $x_0(t)$ and covariance $P(t)$, satisfying the finite ODE system
\begin{equation}\label{eq:linear_cov}
    \dot{x}_0 = A\,x_0, \qquad \dot{P} = A P + P A^\top + \Sigma_0.
\end{equation}
No approximation is needed in this case; instead, a finite set of scalars, the $n$ components of the mean and the $n(n+1)/2$ independent elements of the covariance fully encode the PDF at all times. In the present framework, this corresponds to taking $F(x,t) = \Phi(t)\bigl(x - x_0(t)\bigr)$, where $\Phi(t)$ is the State Transition Matrix (STM) satisfying $\dot{\Phi} = A\Phi$, $\Phi(t_0) = I$. With this choice $J_F = \Phi$ and all higher-order terms vanish, so the system~\eqref{eq:dFdt}--\eqref{eq:dQdt} reduces exactly to~\eqref{eq:linear_cov}. Taylor Map Diffusion generalizes this structure directly to nonlinear dynamics: the nominal trajectory $x_0$ and STM $\Phi$ are generalized to a full nonlinear Taylor map $F$, capturing non-Gaussian deformations through its higher-order coefficients, and the precision matrix $Q$ plays the role of the inverse covariance, governing the spread of the distribution. The result is an ODE system of the same type, finite-dimensional and integrable by any standard solver, that characterizes a non-Gaussian density in the way that mean and covariance characterize a Gaussian one.

\subsection{Physical Interpretation}

Equation~\eqref{eq:dFdt} governs the evolution of the \emph{shape} of the distribution. Its first term ($-J_F f$) is pure advection: the contours of the PDF are transported along the deterministic flow, exactly as in the classical Liouville equation. The remaining terms represent the diffusion correction: process noise smoothly deforms the map $F$, stretching the distribution in directions dictated by the local geometry of the flow and the structure of the noise.

Equation~\eqref{eq:dQdt} governs the evolution of the \emph{concentration} of the distribution. The precision matrix $Q$ decreases over time (the distribution broadens) at a rate determined by the Hessian of the map, that is, by the local curvature of the nonlinear flow. This is a fundamentally different mechanism from simple covariance growth: the rate at which uncertainty spreads depends on the nonlinear structure of the dynamics through $J_F^{-1}$ and $M_i$.

For conservative dynamics ($\nabla\cdot f = 0$), which includes Keplerian motion, $\mu$ remains identically constant (one can verify directly from~\eqref{eq:dmudt} that $\mu = 1$ with $\nabla\mu = 0$ and $\nabla^2\mu = 0$ satisfies the equation when $\nabla\cdot f = 0$). The evolution equation for $\mu$ therefore drops out entirely, and the system reduces to the two-equation conservative form:
\begin{align}
    \frac{\partial F}{\partial t} &= -J_F\,f + \frac{1}{2}\!\left(F\sum_i\alpha_i + Q^{-1}\!\sum_i D_i + \Delta - \Gamma\right), \label{eq:dFdt_cons}\\[4pt]
    \frac{dQ}{dt} &= \frac{1}{2}\sum_i J_F^{-\top}\,J_{\Lambda_i}^\top Q\,J_{\Lambda_i}\,J_F^{-1}. \label{eq:dQdt_cons}
\end{align}
This is the system integrated in the Keplerian orbit application of Section~\ref{sec:keplerian}.

\subsection{Structural Preservation and Tail Modeling}

The preservation of the structural form~\eqref{eq:ansatz} under the full FPE dynamics has a direct practical consequence: the tails of the distribution are determined by the same finite set of parameters $(F,Q)$ that describe the bulk. Once the augmented ODE system is integrated, the density $p(x,t)$ can be evaluated at any point in state space, including at $5\sigma$ or $10\sigma$ deviations from the mode, at negligible additional cost. The non-Gaussian shape of the tails is encoded in the nonlinear map $F$, while their rate of decay is governed by the evolved precision matrix $Q$: both are propagated self-consistently by the coupled dynamics.

This stands in contrast to moment-based methods (which truncate at order 2 or 4 and cannot reconstruct tail behavior), Gaussian mixture models (which require an increasing number of components to capture non-Gaussian tails), and Monte Carlo methods (which need exponentially more samples to resolve lower tail probabilities). For conjunction assessment, where the decision-relevant quantity is a collision probability on the order of $10^{-5}$, having a closed-form expression for $p(x,t)$ that is structurally consistent from the mode to the tails represents a significant operational advantage.

% ============================================================
\section{Application: Keplerian Orbit}\label{sec:keplerian}
% ============================================================

\subsection{Problem Setup}

We apply the Taylor Diffusion algorithm to a spacecraft in an eccentric two-body orbit. The orbit is assumed to be planar, so the full state is $X = [x,\,y,\,v_x,\,v_y]^\top \in \mathbb{R}^4$. The equations of motion are expressed in non-dimensional units (gravitational parameter $\mu_g = 1$, reference length and time chosen accordingly), giving the dynamics:
\begin{equation}
    f(X) = \begin{pmatrix} v_x \\ v_y \\ -x/r^3 \\ -y/r^3 \end{pmatrix}, \qquad r = \sqrt{x^2+y^2}.
\end{equation}
The initial condition is $X_0 = [1,\,0,\,0,\,0.8]^\top$ (apoapsis, sub-circular velocity, yielding an eccentric orbit with semi-major axis $a \approx 0.74$ and period $T \approx 3.96$). The initial PDF is an isotropic Gaussian with $\sigma_{\text{init}} = 10^{-4}$ in all components. Additive process noise acts on velocity only: $\Sigma_0 = \mathrm{diag}(0,\,0,\,\sigma_w^2,\,\sigma_w^2)$ with $\sigma_w = 10^{-4}$. The propagation time is $t_f = 12\pi \approx 37.7$, corresponding to approximately 9.5 complete orbits, a demanding test case in which the spacecraft traverses the strongly nonlinear periapsis region repeatedly, amplifying non-Gaussian distortions with each pass.

\subsection{Implementation}

The map $F$ is represented through its Taylor coefficients up to second order: the nominal trajectory $x_0(t)$, the Jacobian $J(t) \in \mathbb{R}^{4\times 4}$, and the Hessian tensor $H(t) \in \mathbb{R}^{4\times 4\times 4}$. This quadratic truncation captures the leading nonlinear distortion of the PDF; higher-order representations would extend the spatial range of accuracy at the cost of a larger augmented state. Together with $Q(t) \in \mathbb{R}^{4\times 4}$, these form the augmented state $\{x_0, J, H, Q\}$, of total dimension $n + n^2 + n^3 + n(n+1)/2 = 94$ scalar variables for $n=4$.

The implementation leverages the key structural property that all quantities needed to advance the augmented state can be extracted from the map derivative function $\dot{F}(\delta x)$ evaluated \emph{only at $\delta x = 0$}. In particular, let $V_0 = \dot{F}(0)$, $\dot{J} = \partial_{\delta x}\dot{F}\big|_0$, and $\dot{H} = \partial^2_{\delta x}\dot{F}\big|_0$; these three objects fully determine how the Taylor coefficients evolve. The nominal trajectory $x_0$ is advanced by the condition $F(x_0, t) = 0$, which requires $\dot{x}_0 = -J^{-1} V_0$. Since $\dot{F}$ depends on $\delta x$ only through the dynamics $f(x_0 + \delta x)$ and the polynomial map itself, all derivatives are computed automatically via forward-mode automatic differentiation\footnote{We utilize JAX~\cite{jax2018github} for all automatic differentiation and hardware acceleration.}, eliminating the need to derive the (prohibitively complex) 4D analytical expressions by hand.

Since Keplerian dynamics are conservative ($\nabla\cdot f = 0$), the volume parameter $\mu = 1$ is constant and does not require integration. The complete procedure is summarized in Algorithm~\ref{alg:taylor_diffusion}.

\begin{algorithm}[t]
\caption{Taylor Diffusion: augmented ODE integration}\label{alg:taylor_diffusion}
\begin{algorithmic}[1]
\Require Initial state $X_0$, initial covariance $P_0$, process noise $\Sigma_0$, final time $t_f$, steps $N$
\Ensure Augmented state $(x_{0,f},\, J_f,\, H_f,\, Q_f)$ encoding the PDF at $t_f$
\State \textbf{Initialize:} $x_0 \leftarrow X_0$,\; $J \leftarrow I_n$,\; $H \leftarrow 0$,\; $Q \leftarrow \tfrac{1}{2}P_0^{-1}$
\For{$k = 0,\ldots,N-1$} \Comment{RK4 time integration, $\Delta t = t_f/N$}
    \State \textbf{Compute $\dot{Q}$:} $\displaystyle\dot{Q} \leftarrow \tfrac{1}{2}\sum_i \Sigma_{0,i}\,J^{-\top} H_{\alpha i \beta}^\top Q\, H_{\alpha i \beta}\,J^{-1}$
    \State \textbf{Define} $\dot{F}(\delta x)$ per Eq.~\eqref{eq:dFdt_cons}, using $F(\delta x) = J\delta x + \tfrac{1}{2}H(\delta x, \delta x)$
    \State \textbf{Evaluate at $\delta x = 0$ (via automatic differentiation):}
    \begin{align*}
        V_0 &\leftarrow \dot{F}(0), \quad
        \dot{J} \leftarrow \partial_{\delta x}\dot{F}\big|_0 + H\cdot\dot{x}_0, \quad
        \dot{H} \leftarrow \partial^2_{\delta x}\dot{F}\big|_0
    \end{align*}
    \State \textbf{Zero tracking} (enforce $F(x_0)=0$): $\dot{x}_0 \leftarrow -J^{-1}V_0$
    \State \textbf{Propagation:} $(x_0,\, J,\, H,\, Q) \leftarrow \mathrm{RK4\text{-}step}(\dot{x}_0,\, \dot{J},\, \dot{H},\, \dot{Q},\, \Delta t)$
\EndFor
\State \Return $(x_{0,f},\, J_f,\, H_f,\, Q_f)$
\end{algorithmic}
\end{algorithm}

\subsection{Validation and Results}

We validate the Taylor Diffusion PDF against a Monte Carlo ensemble of 400000 stochastic trajectories, each propagated via RK4 with Euler-Maruyama noise injection at 2000 time steps. The analytical PDF, encoded by $(J_f, H_f, Q_f)$ at time $t_f$, is sampled by drawing from $\mathcal{N}(0,\,(2Q_f)^{-1})$ in the map's domain and applying the quadratic forward map.

A key point of comparison is computational cost. The Taylor Diffusion solver integrates an augmented ODE system of dimension $n + n^2 + n^3 + n(n+1)/2$ (nominal trajectory, Jacobian, Hessian, and precision matrix), for $n=4$, this amounts to 94 scalar ODEs\footnote{While the 4D state requires integrating 94 scalar ODEs, extending this framework to a full 6D orbital state yields an augmented system of 279 ODEs. This computational load remains trivially fast for modern numerical integrators, preserving the method's stark operational advantage over Monte Carlo sampling.}. This single integration, taking less than a second\footnote{The computing times have been measured on a \textit{11th Gen Intel(R) Core(TM) i7-1165G7} CPU.}, produces the complete analytical description of the PDF at the final time. In contrast, the Monte Carlo validation required 400000 independent stochastic trajectory integrations, a computation that takes more than a minute, and still subject to sampling noise.

\begin{figure}[ht]
    \centering
    \includegraphics[width=\textwidth]{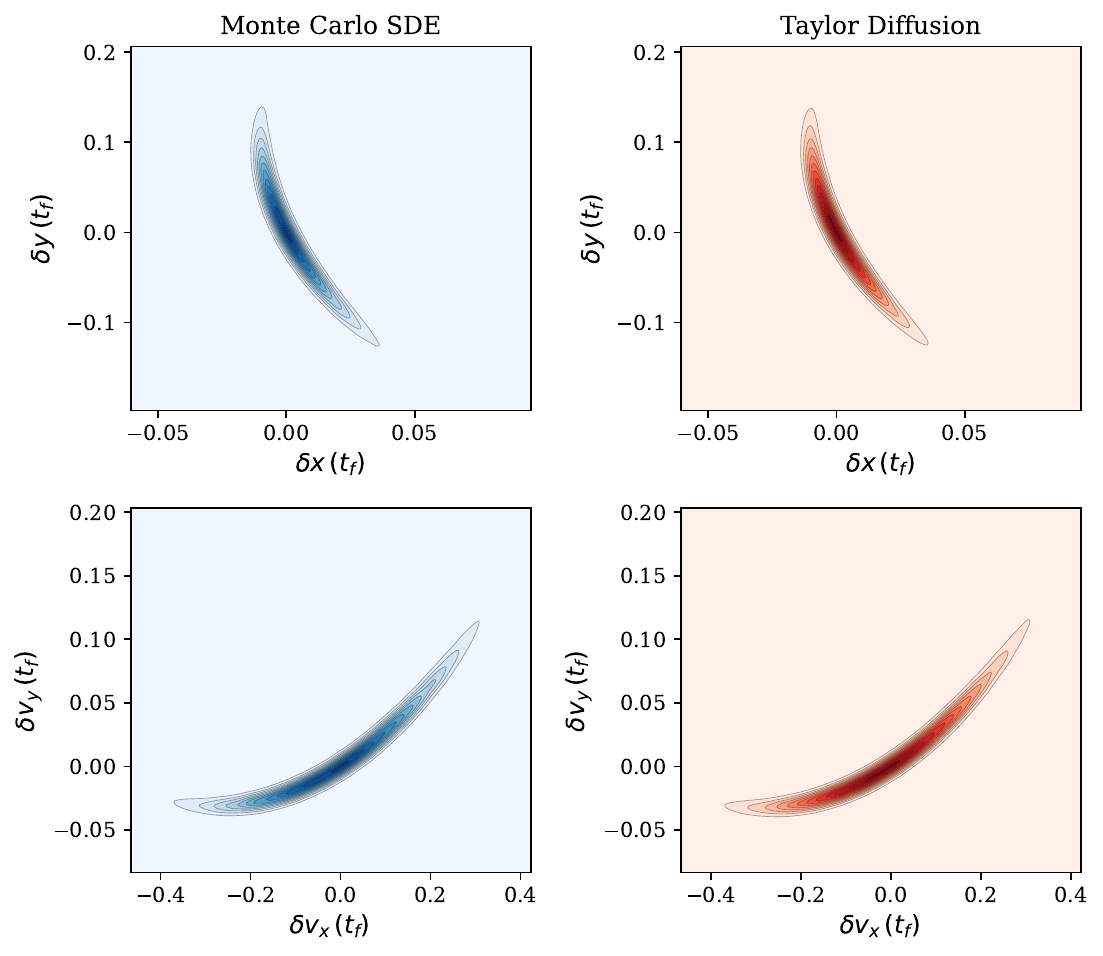}
    \caption{Two-dimensional marginal PDFs at $t_f = 12\pi$ ($\approx$ 9.5 orbits) for the eccentric Keplerian orbit. Left column: Monte Carlo SDE reference (400000 samples). Right column: Taylor Diffusion (second-order map with evolved $Q$). Top row: position subspace $(\delta x, \delta y)$. Bottom row: velocity subspace $(\delta v_x, \delta v_y)$.}
    \label{fig:joint}
\end{figure}

\begin{figure}[ht]
    \centering
    \includegraphics[width=\textwidth]{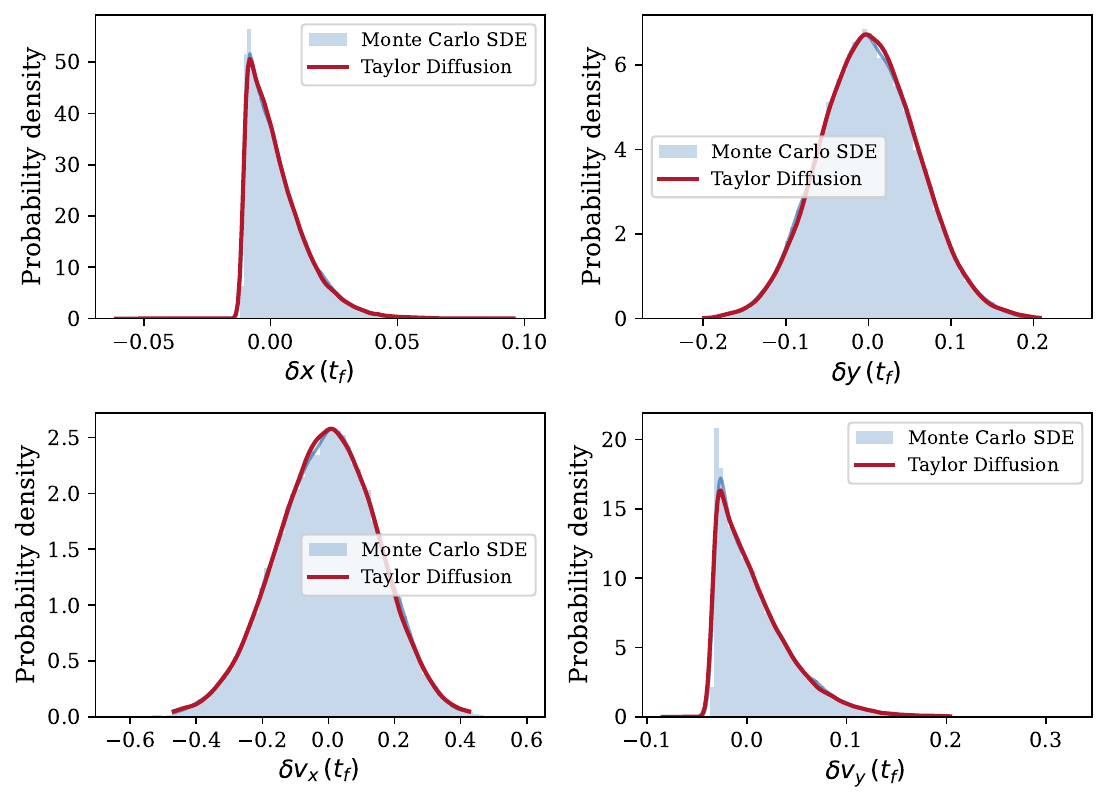}
    \caption{One-dimensional marginal PDFs for each state component. Blue histograms: Monte Carlo SDE. Red curves: Taylor Diffusion. The close agreement across all four components, including the asymmetric tails in $\delta x$ and $\delta v_y$, confirms that the method captures the non-Gaussian structure induced by the nonlinear dynamics and stochastic forcing.}
    \label{fig:marginals}
\end{figure}

The results show close agreement between the Taylor Diffusion and Monte Carlo distributions across both position $(x,y)$ and velocity $(v_x, v_y)$ subspaces, even after 9.5 orbits of propagation. Fig.~\ref{fig:joint} compares the 2D marginal densities: the method correctly captures the non-Gaussian features induced by the $1/r^2$ gravitational nonlinearity, notably the strongly curved, banana-shaped distributions and asymmetric tails, while simultaneously reproducing the broadening due to cumulative stochastic forcing through the evolved precision matrix $Q_f$. Fig.~\ref{fig:marginals} shows the 1D marginal distributions along each state component, demonstrating close correspondence in peak location, width, and asymmetry. The skewed tails visible in $\delta x$ and $\delta v_y$ are characteristic of the nonlinear stretching at periapsis and are faithfully reproduced by the second-order Taylor map.

% ============================================================
\section{Conclusions and Outlook}
% ============================================================

We have presented Taylor Map Diffusion, a framework that provides closed-form, grid-free solutions to the Fokker-Planck equation for nonlinear dynamical systems with additive Gaussian noise. The theoretical foundation is rigorous: the exponential-of-quadratic-form ansatz satisfies the FPE with zero residual under the derived coupled ODEs, with no approximation introduced at the continuous level. In practice, the map $F$ is represented as a finite-order Taylor expansion, making the truncation order the single, transparent control parameter governing the accuracy of the computed density. Even at second order, the method captures the dominant nonlinear distortions of the PDF, including its non-Gaussian tails, while simultaneously and self-consistently accounting for the broadening effect of process noise through the evolved precision matrix $Q$.

Applied to an eccentric Keplerian orbit propagated over 9.5 complete revolutions with continuous stochastic velocity perturbations, the method produces PDFs in close agreement with a 400000-sample Monte Carlo validation, while requiring only the integration of 94 coupled ODEs, with a single trajectory's worth of computation. This disparity in computational cost grows with the number of Monte Carlo samples needed: resolving a $10^{-5}$ tail probability to statistical significance typically requires $10^6$-$10^7$ samples, whereas the Taylor Diffusion solution provides the density analytically at any point in state space, including arbitrarily deep into the tails.

The framework opens several directions for future work. A natural extension is to perturbed orbits with non-conservative forces (atmospheric drag, solar radiation pressure), which require evolving the volume parameter $\mu$ and would demonstrate the method's full generality beyond the conservative case. Higher-order Taylor map expansions would extend accuracy to longer propagation horizons and stronger nonlinearities. Conjunction assessment is a direct target application: the closed-form density enables direct evaluation of collision probabilities in the distribution tails without resorting to sampling.

The present work considers a planar orbit; extension to the full three-dimensional case (six-dimensional state) is straightforward in principle, increasing the augmented ODE system from 94 to 279 scalar equations, still trivially fast for modern integrators. A more significant generalization concerns the choice of state representation. Cartesian coordinates, used here, are natural for the FPE but introduce strong nonlinearities over long arcs. A reformulation in terms of orbital elements, particularly non-singular sets such as the Generalized Equinoctial Orbital Elements (GEqOE)~\cite{Hernando2023}, would reduce the effective nonlinearity of the flow map and thereby improve the accuracy of low-order Taylor expansions over many revolutions, while also simplifying the treatment of near-circular and near-equatorial orbits. The extension of the GEqOE framework to include continuous process noise, currently an open problem, is a natural target for the Taylor Diffusion mechanism developed here.

A further direction of considerable practical interest is nonlinear orbit determination. Many conventional filtering approaches are constrained by Gaussian approximations (e.g., extended and unscented Kalman filters). This formulation degrades significantly when measurement arcs are sparse and propagation times are long, which is exactly the regime where non-Gaussian effects become most pronounced. Taylor Map Diffusion provides a closed-form, non-Gaussian prior that can be updated analytically upon each measurement, offering a principled route to sequential orbit determination that remains tractable even with large temporal gaps between observations. The combination of mathematical rigor, systematic improvability through expansion order, and computational efficiency makes Taylor Map Diffusion a promising tool for next-generation Space Domain Awareness.

% ============================================================
\section*{Acknowledgments}

This work was supported by the Air Force Office of Scientific Research (AFOSR) under Grant No. FA8655-25-1-7012.

% ============================================================
\bibliography{references}

\end{document}